\documentclass[a4paper,11pt]{article}
\usepackage{jheppub} % for details on the use of the package, please
\usepackage[T1]{fontenc} % if needed

\title{}

\author{Marco Bochicchio}
\affiliation[a]{INFN sez. Roma 1\\Piazzale A. Moro 2, Roma, I-00185, Italy}
\affiliation[b]{Scuola Normale Superiore (SNS)\\Piazza dei Cavalieri 7, Pisa, I-56100, Italy}

\emailAdd{marco.bochicchio@roma1.infn.it}

\abstract{We find an asymptotic solution for two-, three- and multi-point correlators of local gauge-invariant operators, in a lower-spin sector of massless large-$N$ $QCD$, in terms of glueball and meson propagators, in such a way that the solution is asymptotic in the ultraviolet to renormalization-group improved perturbation theory, by means of a new purely field-theoretical technique that we call the asymptotically-free bootstrap, based on a recently-proved asymptotic structure theorem for two-point correlators. The asymptotically-free bootstrap provides as well asymptotic $S$-matrix amplitudes in terms of glueball and meson propagators. Remarkably, the asymptotic $S$-matrix depends only on the unknown particle spectrum, but not on the anomalous dimensions, as a consequence of the $LSZ$ reduction formulae. Very many physics consequences follow, both practically and theoretically. 
In fact, the asymptotic solution sets the strongest constraints on any actual solution of large-$N$ $QCD$, and in particular on any string solution.}

\usepackage{braket}
\usepackage{amsmath}
\usepackage{amssymb}
\usepackage{graphicx}
\usepackage{hyperref}
\usepackage{fancyhdr}
\newcommand{\Lambdams}{\Lambda_{\overline{MS}}}

\newcommand{\plms}{\frac{p^2}{\Lambdams^2}}

\def\beq{\begin{equation}}
\def\eeq{\end{equation}}
\def\bea{\begin{eqnarray}}
\def\eea{\end{eqnarray}}
\def\bq{\begin{quote}}
\def\eq{\end{quote}}

\title{An asymptotic solution of large-$N$ $QCD$}
\date{}
\begin{document}
\maketitle
%With large-$N$ Yang-Mills theory we mean Yang-Mills theory with gauge group $SU(N)$ where $\frac{1}{N}$ is used as an expansion %parameter.
%
%In such an expansion diagrams are rearranged according to their topology and at the leading order there are planar graphs.
%This expansion lead to a simplification because the number of planar graphs increases with the number of vertices only %exponentially, while the total number of all graphs increases factorially.
% correspondence
%From the path integral point of view the large-$N$ limit results in the existence of a new saddle point.
%
% 
%\tableofcontents  

%\thispagestyle{empty}

%

\section{Introduction}
\label{s0}
Solving $QCD$ in 't Hooft large-$N$ limit \cite{H1} is a long-standing difficult problem. An easier problem is to find a solution, not exact, but only asymptotic in the ultraviolet ($UV$). 
In a sense this asymptotic solution in the $UV$ already exists: It is ordinary perturbation theory. 
But in fact it is much more interesting an asymptotic solution in the $UV$ written in terms of glueballs and mesons as opposed to gluons and quarks. 
An asymptotic solution of this kind would replace $QCD$ viewed as a theory of gluons and quarks, that are strongly coupled in the infrared in perturbation theory, with a theory of an infinite number of glueballs and mesons, that are weakly coupled at all scales in the large-$N$ limit \cite{H1}. Indeed, at the leading $\frac{1}{N}$ order the two-point connected correlators of gauge invariant operators must be a sum of propagators of free fields \cite{Mig}, involving by the Kallen-Lehmann representation single-particle pure poles, because the interaction associated to three- and multi-point correlators vanishes.  At the next order the interaction arises, but it is parametrically weak in the $\frac{1}{N}$ expansion.  
Recently, the asymptotic structure of two-point correlators of any spin has been explicitly characterized by the asymptotic theorem \cite{MBN} reported below, in 't Hooft limit of any large-$N$ confining asymptotically-free gauge theory massless in perturbation theory, such as
massless $QCD$ (i.e. $QCD$ with massless quarks). 
The asymptotic theorem for the two-point correlators is the basis of a new technique described in this paper, that we call the asymptotically-free bootstrap \footnote{The name derives by the celebrated conformal bootstrap.}, by which we extend the asymptotic theorem to three- and multi-point correlators
and $S$-matrix amplitudes, getting in this way an asymptotic solution of large-$N$ $QCD$ in a sense specified below.  
The asymptotic theorem is based on the Callan-Symanzik equation, plus the Kallen-Lehmann representation, plus the assumption that the theory confines, i.e. technically that the one-particle spectrum for each integer spin $s$ at the leading $\frac{1}{N}$ order is a discrete diverging sequence with asymptotic distribution $ \rho_s(m^2)$. In this introduction we recall the precise statement of the asymptotic theorem, because it is necessary to explain the logic of this paper. 
The connected two-point Euclidean correlator of a local gauge-invariant single-trace operator (or of a fermion bilinear) $\mathcal{O}^{(s)}$ of integer spin $s$ and naive mass dimension $D$ and with anomalous dimension $\gamma_{\mathcal{O}^{(s)}}(g)$,
must factorize asymptotically for large momentum, and at the leading order in the large-$N$ limit, over the following poles and residues (after analytic continuation to Minkowski space-time):
\bea \label{at0}
\langle \mathcal{O}^{(s)}(x) \mathcal{O}^{(s)}(0) \rangle_{conn} 
\sim \sum_{n=1}^{\infty} \frac{1}{(2 \pi)^4} \int  P^{(s)} \big(\frac{p_{A}}{m^{(s)}_n}\big) \frac{m^{(s)2D-4}_n Z_n^{(s)2} \rho_s^{-1}(m^{(s)2}_n)}{p^2+m^{(s)2}_n  } \,e^{ip\cdot x}d^4p \nonumber \\
\eea
where $ P^{(s)} \big( \frac{p_{A}}{m^{(s)}_n} \big)$ is a dimensionless polynomial in the four momentum $p_{A}$ \footnote{ We employ latin letters $A,\cdots$ to denote vector indices, and greek letters $\alpha, \dot \alpha, \cdots$ to denote spinor indices.} that projects on the free propagator of spin $s$ and mass $m^{(s)}_n$ and:
\bea \label{g}
\gamma_{\mathcal{O}^{(s)}}(g)= - \frac{\partial \log Z^{(s)}}{\partial \log \mu}=-\gamma_{0} g^2 + O(g^4)
\eea 
with $Z_n^{(s)}$ the associated renormalization factor computed on shell, i.e. for $p^2=m^{(s)2}_n$:
\bea \label{z}
Z_n^{(s)}\equiv Z^{(s)}(m^{ (s)}_n)= \exp{\int_{g (\mu)}^{g (m^{(s)}_n )} \frac{\gamma_{\mathcal{O}^{(s)}} (g)} {\beta(g)}dg}
\eea
The symbol $\sim$ means always in this paper asymptotic equality in the sense specified momentarily, up to perhaps a constant factor overall. 
The proof of the asymptotic theorem reduces to showing that Eq.(\ref{at0})
matches asymptotically for large momentum, within the universal leading and next-to-leading logarithmic accuracy,
the renormalization-group ($RG$) improved perturbative result implied by the Callan-Symanzik equation. 
An important corollary \cite{MBN} of the asymptotic theorem is that to compute the asymptotic behavior we need not to know explicitly neither the actual spectrum nor the asymptotic spectral distribution, since it cancels by evaluating the sum in Eq.(\ref{at0}) by the integral that occurs as the leading term in Euler-MacLaurin formula. Hence $RG$-improved perturbation theory does not contain in fact spectral information \cite{MBN}, as perhaps expected, and so it does not our asymptotic solution. 
In order to get spectral information it is necessary to lift the asymptotic solution to the actual solution (see conclusions in section \ref{s1}). 
Nevertheless, from a practical point of view, the asymptotic formulae suitably interpreted can be employed also in the infrared, simply substituting the known experimental masses (and in some cases residues, as for $f_{\pi}$ ) of mesons and glueballs, in order to get correlators and $S$-matrix amplitudes that are both factorized over poles of physical particles and are asymptotic to the correct result in the ultraviolet. In this paper we write asymptotic correlators of vector and axial currents, relevant for the light by light scattering amplitude and for the structure of the pion form factor, but we do not discuss at all the physical applications.
This paper is a short communication, detailed proofs will appear elsewhere.    

\section{Asymptotically-free bootstrap for massless large-$N$  $QCD$} \label{s2}

The first step to work out the asymptotically-free bootstrap consists in exploiting the conformal invariance of two- and three-point correlators at the lowest non-trivial order of perturbation theory, together
with the $RG$ corrections implied by the Callan-Symanzik equation, in any asymptotically-free
theory massless in perturbation theory. 
As a consequence, for the connected correlators of a scalar operator $\mathcal{O}$ of naive mass-dimension $D$, $G^{(2)}$ and $G^{(3)}$:
\begin{equation}
\braket{\mathcal{O}(x_1)\mathcal{O}(x_2)}_{\mathit{conn}}=G^{(2)}(x_1-x_2)
\end{equation}
\begin{equation}
 \braket{\mathcal{O}(x_1)\mathcal{O}(x_2) \mathcal{O}(x_3)}_{\mathit{conn}}= G^{(3)}(x_1-x_2, x_2-x_3, x_3-x_1)
\end{equation}
we get the estimates:
\begin{equation}\label{21}
 G^{(2)}(x_1-x_2) \sim  C_2 (1+O(g^2))
\frac{(\frac{g(x_1-x_2)}{g(\mu)})^{\frac{2\gamma_0}{\beta_0}}}{  (x_1- x_2)^{2D}} 
\end{equation}
\begin{equation} \label{31}
G^{(3)}(x_1-x_2, x_2-x_3, x_3-x_1)
 \sim
C_3 (1+O(g^2))\frac{(\frac{g(x_1-x_2)}{g(\mu)})^{\frac{\gamma_0}{\beta_0}}}{(x_1-x_2)^{D}}\frac{(\frac{g(x_2-x_3)}{g(\mu)})^{\frac{\gamma_0}{\beta_0}}}{(x_2-x_3)^{D}}\frac{(\frac{g(x_3-x_1)}{g(\mu)})^{\frac{\gamma_0}{\beta_0}}}{(x_3-x_1)^{D}}
\end{equation}
Eq.(\ref{31}) can be proved by means of the operator product expansion ($OPE$) as well, that allows us to convey more local information than the Callan-Symanzik equation alone, and that essentially reduces the
asymptotic estimates for three-point correlators to two-point correlators. 
In fact, under the assumption that the three-point correlator $\braket{\mathcal{O}(x)\mathcal{O}(0) \mathcal{O}(y)}_{\mathit{conn}}$ does not vanish at lowest order in perturbation theory, i.e. $C_3\neq 0$, we can substitute in the correlator, with asymptotic accuracy as $x$ vanishes, the contribution in the $OPE$ that contains the operator $\mathcal{O}$ itself: $\mathcal{O}(x)\mathcal{O}(0) \sim  C(x) \mathcal{O}(0) + \cdots$,
with $C(x) \sim \frac{(\frac{g(x)}{g(\mu)})^{\frac{\gamma_0}{\beta_0}}}{ x^{D}} $ by the Callan-Symanzik equation for the coefficient functions in the $OPE$.
Hence we get asymptotically for $x_1 \rightarrow x_2$:
\bea \label{OPE}
\braket{\mathcal{O}(x_1)\mathcal{O}(x_2) \mathcal{O}(x_3)}_{\mathit{conn}} && \sim C(x_1-x_2)\braket{\mathcal{O}(x_2)\mathcal{O}(x_3)}_{\mathit{conn}}  \sim C(x_1-x_2) G^{(2)}(x_2-x_3) \nonumber \\
 &&\sim
C(x_1-x_2)C^2(x_2-x_3)
\eea
that coincides with Eq.(\ref{31}). Therefore, because of the symmetric nature of Eq.(\ref{31}) and by Eq.(\ref{OPE}), we get the fundamental result valid for $C_3 \neq 0$:
\bea \label{AS}
G^{(3)}(x_1-x_2, x_2-x_3, x_3-x_1) \sim C(x_1-x_2) C(x_2-x_3) C(x_3-x_1)
\eea
So far so good. Everything that we have discussed is well known, but perhaps Eq.(\ref{AS}), in any asymptotically-free theory massless in perturbation theory.
Now it comes the interesting part. 
The asymptotic theorem extends to the coefficients of the $OPE$ in the scalar case, because they arise from the non-perturbative part involving condensates of the scalar two-point correlator, which the
Kallen-Lehmann representation applies to as well. In particular for $C(x)$ we get:
\bea \label{001}
&&C(x_1-x_2)
\sim \sum_{n=1}^{\infty} \frac{1}{(2 \pi)^4} \int   \frac{m^{D-4}_n Z_n \rho^{-1}(m^{2}_n)}{p^2+m^{2}_n  } \,e^{ip\cdot (x_1-x_2)}d^4p \nonumber \\
&&\sim \sum_{n=1}^{\infty} \frac{1}{(2 \pi)^4} \int   \frac{m^{D-4}_n (\frac{g(m_n)}{g(\mu)})^{\frac{\gamma_0}{\beta_0}} \rho^{-1}(m^{2}_n)}{p^2+m^{2}_n  } \,e^{ip\cdot (x_1-x_2)}d^4p 
 \sim  
\frac{(\frac{g(x_1-x_2)}{g(\mu)})^{\frac{\gamma_0}{\beta_0}}}{  (x_1- x_2)^{D}} 
\eea
The basic idea of the asymptotically-free bootstrap is to substitute the Kallen-Lehmann representation for $C(x)$, Eq.(\ref{001}), in Eq.(\ref{AS}).
Thus explicitly and constructively we get the asymptotic spectral representation of three-point scalar correlators in momentum space:
\bea \label{2}
&&\braket{\mathcal{O}_{D, \gamma_0}(q_1)\mathcal{O}_{D, \gamma_0}(q_2) \mathcal{O}_{D, \gamma_0}(q_3)}_{\mathit{conn}}  \nonumber \\
&&\sim \delta(q_1+q_2+q_3) \sum_{n_1=1}^{\infty}       \sum_{n_2=1}^{\infty}  \sum_{n_3=1}^{\infty} \int   \frac{m^{D-4}_{n_1} Z_{n_1}\rho^{-1}(m^{2}_{n_1})}{p^2+m^{2}_{n_1}  } 
   \frac{m^{D-4}_{n_2} Z_{n_2}\rho^{-1}(m^{2}_{n_2})}{(p+q_2)^2+m^{2}_{n_2}  }   
  \frac{m^{D-4}_{n_3} Z_{n_3}\rho^{-1}(m^{2}_{n_3})}{(p+q_2+q_3)^2+m^{2}_{n_3}  }d^4p \nonumber \\
\eea
But this cannot be the whole story.  
Indeed, while Eq.(\ref{2}) is asymptotic to the correct result in $RG$-improved perturbation theory, it has not the correct pole structure, that is a consequence of the $OPE$:
\bea \label{ope}
&& \lim_{q^2_2\rightarrow \infty} \braket{ \mathcal{O}_{D, \gamma_0}(q_1) \mathcal{O}_{D, \gamma_0}(q_2)\mathcal{O}_{D, \gamma_0}(q_3) }_{\mathit{conn}}  
 \sim \lim_{q^2_2 \rightarrow \infty} \delta(q_1+q_2+q_3)        
   C(q_2)  G^{(2)}(q_3)    \nonumber \\ 
&& \sim \lim_{q_2^2 \rightarrow \infty} \delta(q_1+q_2+q_3) \sum_{n_1=1}^{\infty}            
     \frac{m^{D-4}_{n_1} Z_{n_1}\rho^{-1}(m^{2}_{n_1})}{q_2^2+m^{2}_{n_1}  } \sum_{n_2=1}^{\infty}     \frac{m^{2D-4}_{n_2} Z^2_{n_2}\rho^{-1}(m^{2}_{n_2})}{q_3^2+m^{2}_{n_2}  }  \nonumber\\
     \eea
as it follows Fourier transforming in Eq.(\ref{OPE}) and substituting the Kallen-Lehmann representation. 
Thus while Eq.(\ref{ope}) and Eq.(\ref{2}) have the same large-momentum asymptotics, the symmetric form and the $OPE$ form factorize on different cuts and poles. 
Indeed, $RG$-improved perturbation theory has a non-perturbative ambiguity by multiplicative functions of the external Euclidean momenta, that are asymptotic to $1$ in the ultraviolet. 
We fix asymptotically this ambiguity by requiring that the new improved three-point correlator carries a simple pole for each external momentum $(q_1,q_2,q_3)$ on shell in Minkowski space-time, but without changing its Euclidean asymptotic behavior. 
Hence, the real structure of the Euclidean correlator must be asymptotically:
\bea \label{3}
&& \braket{\mathcal{O}_{D, \gamma_0}(q_1)\mathcal{O}_{D, \gamma_0}(q_2) \mathcal{O}_{D, \gamma_0}(q_3)}_{\mathit{conn}}  
 \sim  \delta(q_1+q_2+q_3)        
   \int \sum_{n_1=1}^{\infty}  \frac{m^{D-4}_{n_1} Z_{n_1}\rho^{-1}(m^{2}_{n_1})}{p^2+m^{2}_{n_1}  } \frac{m_{n_1}^2}{q_2^2+m_{n_1}^2}\nonumber\\
 &&\sum_{n_2=1}^{\infty}    \frac{m^{D-4}_{n_2} Z_{n_2}\rho^{-1}(m^{2}_{n_2})}{(p+q_2)^2+m^{2}_{n_2}  }   \frac{m_{n_2}^2}{q_3^2+m_{n_2}^2}
   \sum_{n_3=1}^{\infty}     \frac{m^{D-4}_{n_3} Z_{n_3}\rho^{-1}(m^{2}_{n_3})}{(p+q_2+q_3)^2+m^{2}_{n_3}  }   \frac{m_{n_3}^2}{q_1^2+m_{n_3}^2} d^4p 
\eea
where we employed: $\lim_{n \rightarrow \infty} \frac{m_n^2}{q^2+m_n^2}=1$. Proceeding by induction in the $OPE$ and employing $x_1 \sim x_2$, we get the asymptotic contribution to the $r$-point scalar correlator:
\bea \label{r}
\braket{ \mathcal{O}(x_1) \mathcal{O}(x_2)    \cdots \mathcal{O}(x_r) }_{\mathit{conn}} \sim   C(x_1-x_2) \cdots C(x_r-x_1) 
\eea 

\section{Asymptotic effective action and $S$-matrix amplitudes} \label{s4}

From Eq.(\ref{3}) it follows the asymptotic effective action in the scalar glueball sector at lower orders, that reproduces the correlators in the $\frac{1}{N}$ expansion by means of the identification
$\mathcal{O}(x)=\sum_n \Phi_n(x)$:
\bea \label{eff1}
&&\Gamma=  \frac{1}{2!} \sum_n \int dq_1 dq_2 \delta(q_1+q_2) m_n^{4-2D} Z_n^{-2}\rho(m^{2}_{n}) \Phi_n(q_1) (q_1^2+m_n^2) \Phi_n(q_2) + \frac{C}{3! N} \int dq_1 dq_2 dq_3       \nonumber \\
&&\delta(q_1+q_2+q_3)
 \int \sum_{n_1=1}^{\infty} m_{n_1}^2 \frac{m^{-D}_{n_1} Z^{-1}_{n_1} \Phi_{n_1}(q_2)}{p^2+m^{2}_{n_1}  } \sum_{n_2=1}^{\infty}   m_{n_2}^2 \frac{m^{-D}_{n_2} Z^{-1}_{n_2}\Phi_{n_2}(q_3)}{(p+q_2)^2+m^{2}_{n_2}  }   
   \sum_{n_3=1}^{\infty}      m_{n_3}^2\frac{m^{-D}_{n_3} Z^{-1}_{n_3}\Phi_{n_3}(q_1)}{(p+q_2+q_3)^2+m^{2}_{n_3}  }dp \nonumber \\
\eea
with $C \sim O(1)$, computable in lowest-order perturbation theory.
Fourier transforming Eq.(\ref{r}) and employing the Kallen-Lehmann representation for $C(x)$, we get the spectral representation of the asymptotic primitive $r$-point scalar vertices in the effective action up to overall normalization:
\bea \label{hl}
 &&\int dq_1 dq_2 \cdots dq_r \delta(q_1+q_2+ \cdots + q_r)        
   \int \sum_{n_1=1}^{\infty} m_{n_1}^2 \frac{m^{-D}_{n_1} Z^{-1}_{n_1} \Phi_{n_1}(q_2)}{p^2+m^{2}_{n_1}  } 
 \sum_{n_2=1}^{\infty}   m_{n_2}^2 \frac{m^{-D}_{n_2} Z^{-1}_{n_2}\Phi_{n_2}(q_3)}{(p+q_2)^2+m^{2}_{n_2}  }   \nonumber \\
&&\cdots   \sum_{n_r=1}^{\infty}      m_{n_r}^2\frac{m^{-D}_{n_r} Z^{-1}_{n_r}\Phi_{n_r}(q_1)}{(p+q_2+ \cdots +q_r)^2+m^{2}_{n_r}  }dp
\eea
The $S$-matrix generating functional follows setting the kinetic term in canonical form by rescaling the fields $\Phi_n$, that is equivalent to dividing by the square root of the residues of the propagators in the $LSZ$ formulae:
\bea \label{can}
&& S=  \frac{1}{2!} \sum_n \int dq_1 dq_2 \delta(q_1+q_2) \Phi_n(q_1) (q_1^2+m_n^2) \Phi_n(q_2) 
+ \frac{C}{3!N} \int dq_1 dq_2 dq_3 \delta(q_1+q_2+q_3)        \nonumber \\
  &&  \int \sum_{n_1=1}^{\infty} \frac{   \rho^{-\frac{1}{2}}(m^2_{n_1})\Phi_{n_1}(q_2)}{p^2+m^{2}_{n_1}  } 
\sum_{n_2=1}^{\infty}   \frac{ \rho^{-\frac{1}{2}}(m^{2}_{n_2})\Phi_{n_2}(q_3)}{(p+q_2)^2+m^{2}_{n_2}  }   
   \sum_{n_3=1}^{\infty}  \frac{ \rho^{-\frac{1}{2}}(m^2_{n_3})\Phi_{n_3}(q_1)}{(p+q_2+q_3)^2+m^{2}_{n_3}  }dp 
\eea
and analogously for the primitive $r$-point vertices. Remarkably, the dependence on the naive dimension and anomalous dimension has disappeared, because the $S$-matrix cannot depend on the choice of the interpolating field for the same asymptotic state.The asymptotic interaction in the scalar sector is generated by an infinite number of vertices that look like one-loop diagrams in a $\Phi^3$ field theory up to perhaps normalization, one for each order of the $\frac{1}{N}$ expansion. If we consider only power-counting, because of the explicit factor of $\rho_0^{-\frac{1}{2}}$ in Eq.(\ref{can}), that has dimension of $\Lambda_{QCD}$, the $S$-matrix in the scalar sector behaves in the $UV$ 
as in a super-renormalizable field theory. But the actual dependence of the $S$-matrix amplitudes on the masses of the asymptotic states is very sensitive, because of the factors of $\rho_0^{-\frac{1}{2}}$ for each external line, to the rate of grow with the mass of the spectral distribution, that includes the multiplicities, while the factor of $\rho_0^{-1}$ cancels after summing on intermediate states in internal propagators. 
In string theories defined on space-time curved in some extra dimensions it might not be impossible to reproduce such asymptotic $S$-matrix,
but no presently known string theory, in particular based on the $AdS$ String/Gauge Fields correspondence, reproduces the asymptotics of correlators implied by Eq.(\ref{eff1}) (see section \ref{s1}). 
The simplest three-point correlators next to the scalar ones involve the flavor chiral $R=+,-$, vector $R=V$, and axial $R=A$ currents, $j^{a}_{R \alpha \dot \beta}$, built by means of fermion bilinears in $QCD$, with $a$ a flavor index in the adjoint representation of
$SU(N_f)$, or $a=0$ for the identity representation. The generating functional of spin-$1$ correlators in spinor notation, by the identification $j^{a\alpha_1 \dot \beta_1}_{R }(x)=\sum_n \Phi^{a \alpha_1 \dot \beta_1}_{Rn}$, reads:
\bea \label{effspin1}
&&\Gamma_{1R} = \frac{1}{2} \sum_n \int dq_1 dq_2 \delta(q_1+q_2) m_{Rn}^{-2} Z_{Rn}^{-2}  \rho_{1R}(m^2_{Rn})  
\Phi^{a \alpha \dot \beta}_{Rn }(q_1) \big( q_1^2+ m_{Rn}^2  \big) \Phi^a_{ \alpha \dot \beta Rn}(q_2) \nonumber \\
&&+ \frac{C}{3 \sqrt N} \int dq_1 dq_2 dq_3 \delta(q_1+q_2+q_3)   Tr^{(R)}(a,b,c)  
 \int \sum_{n_1=1}^{\infty}  \frac{p_{\alpha_1 \dot \beta_2} m^{-2}_{Rn_1} z^{-1}_{Rn_1} \Phi^{a \alpha_1 \dot \beta_1}_{Rn_1}(q_2)}{p^2+m^{2}_{Rn _1}  } \nonumber \\
&& \sum_{n_2=1}^{\infty}  \frac{(p+q_2)_{\alpha_2 \dot \beta_3} m^{-2}_{Rn_2} z^{-1}_{Rn_2} \Phi^{b \alpha_2 \dot \beta_2}_{Rn_2}(q_3)}{(p+q_2)^2+m^{2}_{Rn _2}  }      
 \sum_{n_3=1}^{\infty}  \frac{(p+q_2+q_3)_{\alpha_3 \dot \beta_1} m^{-2}_{Rn_3} z^{-1}_{Rn_3} \Phi^{c \alpha_3 \dot \beta_3}_{Rn_3}(q_1)}{(p+q_2+q_3)^2+m^{2}_{Rn _3}  } dp  \nonumber \\
\eea
with $\partial_{\alpha \dot \beta}  \Phi^{a \alpha \dot \beta}_ {Rn}(x)=0$ on shell, and to match $RG$-improved perturbation theory: $ \lim_{n \rightarrow \infty} Z_{Rn} =1 $, $\sum_n z_{Rn} m_{Rn}^{-2}  \rho^{-1}_{1}(m^2_{Rn}) \sim 1 $, $Tr^{(+)}(a,b,c)= Tr(T^aT^bT^c)$, $Tr^{(-)}(a,b,c)= - Tr(T^cT^bT^a)$, $ Tr^{(V)}(a,b,c) \sim f^{abc} $, $ Tr^{(A)}(a,b,c) \sim d^{abc} $. The factors of $ p_{\alpha_1 \dot \beta_2} m^{-1}_{Rn_1}$ spoil the super-renormalizability of the $S$-matrix, obtained setting $\Gamma_{1R}$ in canonical form, because now the effective coupling is dimensionless $\rho^{-\frac{1}{2}}_{1R}  m^{-1}_{Rn} $. By power counting, in the spin-$1$ sector for mesons large-$N$ $QCD$ is renormalizable but not super-renormalizable, yet all the divergences must be reabsorbed in a redefinition of $\Lambda_{QCD}$.

\section{ Conclusions and  outlook} \label{s1}
The main limitation of the asymptotic solution is that it does not provide spectral information. 
But first and foremost, the intrinsic interest of the asymptotic solution is that it furnishes a concrete guide to find out an actual solution, possibly only for the spectrum and the $S$-matrix amplitudes, by other methods,
that may be of field-theoretical or of string-theoretical nature. Moreover, for approximate solutions, the asymptotic solution provides a quantitative measure of how good or bad the approximation is. 
For example, employing the asymptotic theorem for two-point correlators \cite{MBN} or directly resumming the leading logarithms of perturbation theory \cite{MBM}, it has become apparent that all the present proposal for the scalar or pseudoscalar glueball propagators in confining asymptotically-free $QCD$-like theories based on the $AdS$ String/Gauge Fields correspondence disagree by powers of logarithms \cite{MBM,MBN} with the asymptotic solution. 
Indeed, by the asymptotic theorem the asymptotic behavior of the scalar glueball propagator, the correlator that controls the mass gap in large-$N$ $YM$, reads in any asymptotically-free gauge theory massless in perturbation theory \cite{MBN,MBM}:
\begin{align}\label{eqn:corr_scalare_inizio}
&\int\langle Tr{F_{}^2}(x) Tr{F}_{}^2(0)\rangle_{conn}e^{ip\cdot x}d^4x 
\sim  p^4\Biggl[\frac{1}{\beta_0\log\frac{p^2}{\Lambdams^2}}\Biggl(1-\frac{\beta_1}{\beta_0^2}\frac{\log\log\frac{p^2}{\Lambdams^2}}{\log\frac{p^2}{\Lambdams^2}}\Biggr)+O\biggl(\frac{1}{\log^2\plms}\biggr)\Biggr]
\end{align} 
up to contact terms, while all the glueball correlators presently computed in the literature on the basis of the $AdS$ String/Gauge Fields correspondence behave as $p^4 \log^n (\frac{p^2}{\mu^2})$, with $n=1$ in the Hard Wall and Soft Wall models, and $n=3$ in the Klebanov-Strassler cascading $\mathcal{N}$ $=1$ $SUSY$ gauge theory, despite in the latter the asymptotically-free $NSVZ$ beta function is correctly reproduced in the supergravity approximation. 
The aforementioned asymptotic disagreement \cite{MBN,MBM} implies that for an infinite number of poles and/or residues the large-$N$ glueball propagator on the string side of the would-be correspondence disagrees with the actual propagator of the asymptotically-free $QCD$-like theory on the gauge side. 
This is unsurprising, since the stringy gravity side of the correspondence is in fact strongly coupled in the $UV$, and therefore it cannot describe the $UV$ of any confining asymptotically-free gauge theory. Thus there is no reason that the needed spectral information be correctly encoded in such class of strongly-coupled $AdS$-based theories.
However, we expect that the $QCD$ string is singled out as the unique string theory that is asymptotic to the asymptotic solution for the $S$-matrix, with the asymptotic states labelled by the spectrum generating algebra of $QCD$.

\end{document}